\documentclass[12pt]{iopart}

\newcommand{\lesssim}{\lower.5ex\hbox{$\; \buildrel < \over\sim \;$}}
\newcommand{\gtrsim}{\lower.5ex\hbox{$\; \buildrel > \over\sim \;$}}

\newcommand{\apj}{{\it Astrophys. J.}}

\usepackage{graphicx}
\begin{document}

\title[High energy radiations from GRBs]{Ultra-high energy 
cosmic rays, cascade $\gamma$-rays, and high-energy neutrinos from gamma-ray bursts}

\author{C D Dermer$^1$ \& A Atoyan$^2$}

\address{$^1$Space Science Division Code 7653 4555 Overlook Ave. 
SW Washington DC 20375 USA}
\ead{dermer@gamma.nrl.navy.mil}
\address{$^2$Centre de Recherches Math\'ematiques 
Universit\'e de Montr\'eal Montr\'eal Canada H3C 3J7}
\ead{atoyan@crm.umontreal.ca}

\begin{abstract}
Gamma-ray bursts (GRBs) are sources of energetic, highly variable
fluxes of $\gamma$ rays, which demonstrates that they are powerful
particle accelerators. Besides relativistic electrons, GRBs should
also accelerate high-energy hadrons, some of which could escape 
cooling to produce
ultra-high energy cosmic rays (UHECRs).  Acceleration of high-energy
hadrons in GRB blast waves will be established if high-energy
neutrinos produced through photopion interactions in the blast wave
are detected from GRBs.  Limitations on the energy in
nonthermal hadrons and the number of expected neutrinos are imposed by
the fluxes from pair-photon cascades initiated in the same processes
that produce neutrinos.  Only the most powerful bursts at fluence
levels $\gtrsim 3\times 10^{-4}\,\rm erg \ cm^{-2}$ offer a realistic
prospect for detection of $\gg$ TeV neutrinos.  Detection of high-energy
neutrinos is likely if GRB blast waves have large baryon loads and
Doppler factors $\lesssim 200$. Cascade $\gamma$ rays will accompany neutrino
production and might already have been detected as anomalous emission
components in the spectra of some GRBs.  Prospects for detection of GRBs in the
Milky Way are also considered.
\end{abstract}

\maketitle

\section{Introduction}

There is general consensus that acceleration of cosmic rays by
supernova remnants (SNRs) is the main source of galactic cosmic
rays at energies below $\sim 100$ TeV (e.\ g., \cite{drury}). It is also
generally thought that all cosmic rays with energies up to at least
the second knee in the cosmic-ray spectrum at $E_2\sim 3\times
10^{17}\,\rm eV $ (e.g. \cite{som01}) or even up to the ankle at
$E_{\rm ank} \simeq 3\times 10^{18}$ eV are produced in our Galaxy
(see, e.g., \cite{nw00}).  Meanwhile, cosmic-ray
acceleration to energies significantly exceeding $\approx 10^{14}$ eV
with the conventional mechanism of nonrelativistic first-order shock
acceleration by SNRs from typical (Type Ia and II) supernovae (SNe) is
problematic \cite{lc83}. The origin of the knee in the cosmic-ray
spectrum in the form of a spectral-index break in the power-law
all-particle spectrum by $\approx 0.3$ units at $E_1 \simeq 3\times
10^{15}\,\rm eV$, accompanied by a change in the cosmic-ray
composition, seems to suggest a new contribution to cosmic rays at
these energies.

GRBs have been proposed as effective accelerators of cosmic rays in
the universe \cite{vie95} and as probable sources of cosmic rays up to
ultra-high energies in our Galaxy \cite{der02}.  Cosmic rays with
energies below $\sim 100$ TeV are produced in the conventional
scenario of continuous injection due to nonrelativistic shock
acceleration by SNRs formed in all types of SNe, with subsequent
modification of the source spectrum through energy-dependent
propagation (see \cite{drury,ber90}).  GRBs in our own Galaxy are very
rare and probably take place at a rate $\lesssim 10^{-4}$
yr$^{-1}$. Our best opportunity to establish cosmic-ray acceleration
in GRB blast waves is by looking for hadronic emission signatures in
the spectra of GRBs, or characteristic signatures from remnants of
past GRBs. Conclusive evidence for cosmic-ray acceleration in GRBs
will come from the detection of high-energy neutrinos.

In the relativistic blast-wave model for GRBs, photohadronic production 
is the most important mechanism for producing high-energy neutrinos. In Section 2,
the photohadron process is described and an approximation used for
computing the spectra of secondaries formed in photohadronic processes, 
which is also useful for analytical studies,
is given. Calculations of secondary neutrino and cascade gamma-ray
production are presented in Section 3, and the requirements for
detection of high-energy neutrinos from GRBs are derived. Models for
UHECRs from GRBs are described in Section 4. A discussion of
signatures of GRBs in the Galaxy is found in Section 5, where we briefly
consider recent reports that GRBs favor sites of low metallicity.  Summary and
conclusions are given in Section 6.

\section{Photohadronic Processes}

If high-energy hadrons are accelerated by GRB blast waves, then
photohadronic processes, which require the presence of target photons,
are the most important hadronic energy-loss mechanisms.  The target
photon field is simply the highly variable radiation formed in the GRB
blast wave that is detected as the GRB.  Secondary nuclear production,
by contrast, requires a large target particle density that would make
the GRB energetics untenable \cite{ad03}.

The most important photohadronic process for hadronic energy losses in
GRB blast waves is the photopion reaction, which can be written as
$p+\gamma \rightarrow N + \pi$, where $N$ stands for a proton or
neutron. Another important photohadronic process is photopair
production, which can be written as $p+\gamma \rightarrow p + e^{+} +
e^{-}$.  A large fraction of the initial proton energy is lost in a
photopion reaction, so only a few scatterings are required for the
proton to lose most of its initial energy. By contrast, only a small
fraction of the initial proton energy is lost in photopair production,
so hundreds of scatterings are needed for a proton to lose most of its
initial energy through this process. Although the photopair reaction
is not so important for hadronic energy losses in GRB blast waves, it
can play an important role in the evolution of the UHECR spectrum
throughout intergalactic space \cite{bgg04}.

The photopion process $p + \gamma \rightarrow N + \pi$ has a threshold
photon energy $\epsilon_{th} = m_\pi + m_\pi^2/2 m_p \cong 150$ MeV,
and its cross section is shown in Fig.\ 1.  Four separate
contributions to the total photopion cross section \cite{muc99} are
shown, namely resonance production involving, for example, the
$\Delta^+(1232)$ resonance; direct production, which refers to
residual, nonresonant contributions to direct two-body channels;
multi-pion production; and diffractive scattering.

\begin{figure}[t]
%\vskip-0.8in
\center
\includegraphics[scale=0.4]{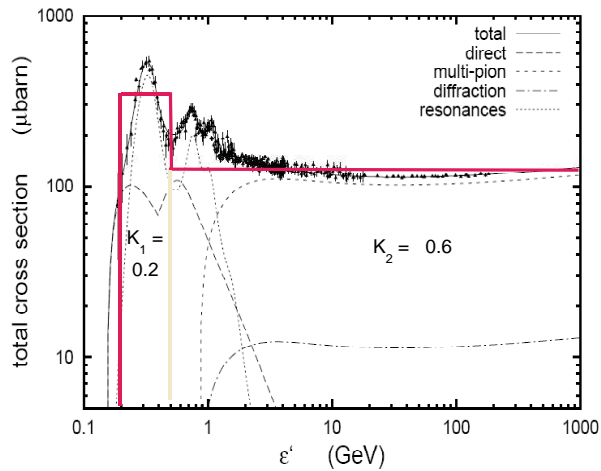}
%\vskip-0.9in
\caption{The total $p\gamma$ photomeson cross section 
as a function of the photon's energy in the proton rest frame
\cite{muc99} ($1\mu$barn = $10^{-30}$ cm$^{2}$; data from Ref.\
\cite{bal88} and references therein). The contributions of baryon
resonances, the direct single-pion channel, diffractive scattering,
and multipion production are shown separately. Also shown is the two
step-function approximation used in calculations.}
\label{f9-1}
\end{figure}

A useful approximation \cite{ad03} to the photopion process is to
treat it as the sum of two channels, namely the {\it single-pion
resonance channel}, where the proton loses 20\% of its energy on
average through the reaction $$ p+\gamma \rightarrow \Delta^+
\rightarrow\cases{ p+\pi^0 \rightarrow p + 2\gamma &$~$\cr\cr n+\pi^+
\rightarrow n + e^+ +3\nu &$~$\cr}\;,$$ 
and a {\it multi-pion channel}, where the proton loses on average 60\% of
its initial energy. The latter channel is assumed to be equally
divided into secondary $\pi^0,\pi^+,$ and $\pi^-$ particles.  In the
$p+\gamma \rightarrow \Delta^+ \rightarrow p+\pi^0$ channel, the
neutral pion decays into two $\gamma$ rays with $\approx 10$\% of the
energy of the initial proton. Following production, the $\gamma$ rays
will usually convert
into $e^+$-$e^-$ pairs through the
$\gamma\gamma \rightarrow e^+$-$e^-$ absorption processes, initiating
an electromagnetic synchrotron/Compton/pair-production cascade.

In the $p+\gamma \rightarrow \Delta^+ \rightarrow n+\pi^+$ channel,
the decay of the charged pion produces three neutrinos and a positron
in the reaction chain $$\pi^+ \rightarrow \mu^+ + \nu_\mu\;,\;{\rm~
followed ~by ~the ~decay } \;\; \mu^+ \rightarrow e^+ + \nu_e + \bar
\nu_\mu.$$ Unless the neutron first interacts with a photon through
the photohadronic process and is converted into a proton, it decays
with a mean lifetime $t_n\cong 886$ s \cite{hag02} in the rest frame through the
$\beta$-decay reaction $$n \rightarrow p+e^-+\bar \nu_e\;.$$ The
$\beta$-decay electron and neutrino have energies $\approx 1$ MeV in
the neutron's rest frame.  In a single $p\gamma$ interaction leading
to the production of a single $\pi^+$, four $\nu$ and two leptons are
therefore formed, with one of the neutrinos and one of the leptons
having $\approx 50\times$ less energy than the others.

The two-step function approximation \cite{ad03,da03} for the photopion
cross section is given by
\begin{equation}
\sigma_{p\gamma}(\epsilon_r) =\cases{340~\mu{\rm b},
&$\epsilon_{th} = 390\leq\epsilon_r \leq 980$\cr\cr
	120~\mu{\rm b},&$\epsilon_r > 980$\cr}\;
\label{sigmaK}
\end{equation}
and inelasticity
\begin{equation}
K_{p\gamma}(\epsilon_r ) =\cases{0.2,&$390\leq\epsilon_r \leq 980$\cr\cr
	0.6,&$\epsilon_r > 980$\cr}\;,
\label{KK1}
\end{equation}
as shown in Fig.\ 1.  Hence
\begin{equation}
\sigma_{p\gamma}(\epsilon_r) K_{p\gamma}
(\epsilon_r) \equiv \sigma_1K_1\cong 70
H(\epsilon_r - 390)~\mu{\rm b}.
\label{sigmaKphipi}
\end{equation}

In this approximation, the lower-energy step function approximates the
$\Delta(1232)$ resonance production process, and the higher-energy
step function approximates the multi-photon production process. For
the $\Delta$ resonance, isospin statistics imply \cite{lm70} a
charge-changing ratio of 1/3 for single resonance production. Thus the
probability of the processes $p\gamma \rightarrow n\pi^+$ and $p\gamma
\rightarrow p\pi^0$ is in the ratio $1:2$.  In the multi-pion limit,
the production of $\pi^+$, $\pi^-$ and $\pi^0$ approaches $1:1:1$.

\section{Neutrinos and Cascade Gamma Rays}

Much effort has been devoted to understanding radiative signatures of
leptons accelerated in GRB blast waves (e.g., \cite{mes02}), and this approach can also be
applied to hadronic acceleration \cite{bd}.  Consider a GRB blast wave with
Lorentz factor $\Gamma$.  Relativistic protons or hadrons are assumed
to be injected in the comoving frame of the blast wave with a number
spectrum $\propto E_p^{-s}$ at comoving proton energies $E_p > \Gamma
$ GeV up to a maximum proton energy determined by the condition that
the particle Larmor radius is smaller than both the size scale of the
emitting region and the photomeson energy-loss length. The injection
index $s \approx 2.2$, as suggested by relativistic shock acceleration
and particle injection determined by analyses of GRB afterglows \cite{gak}.

The total photon fluence $\Phi = \int_{-\infty}^\infty dt \int_0^\infty dE\;E\phi(E,t)$ 
of a GRB, where $\phi(E,t)$ is the differential photon number flux, must be at the level of $\Phi\gtrsim
10^{-4}$ ergs cm$^{-2}$ to be detected with a km-scale neutrino
telescope such as IceCube, as we now show. 
For the detection of $N_{\nu_\mu}$ muon
neutrinos, the best sensitivity of a neutrino telescope for detecting
a spectrum of neutrinos that is falling $\propto \epsilon_\nu^{-p}$
with $p> 1$ is near 100 TeV $\approx 160$ ergs. This is because of the
linear increase of the
detection probability $P_{\nu_\mu}(\epsilon_\nu ) \approx 10^{-4} (\epsilon_\nu/$100 TeV) at
$\epsilon_\nu \gtrsim 1$ TeV \cite{ghs95}, and the increasingly large
cosmic-ray induced neutrino background at $\epsilon_\nu \ll 100$ TeV.  The detection of
$N_{\nu_\mu}$ neutrinos therefore requires that the neutrino fluence
$\Phi_{\nu_\mu} \simeq (160
{\rm~ergs})N_{\nu_\mu}/[P_{\nu_\mu}(100~{\rm TeV})\cdot 10^{10}$ cm$^2$] $\approx 10^{-4}
N_{\nu_\mu}$ ergs cm$^{-2}$. Because the differential neutrino fluence 
$E^2\phi_\nu(E) = E^2\int_{-\infty}^\infty dt\; \phi_\nu (E,t)$, where $\phi_\nu (E)$ 
is the differential neutrino number flux, will be
spread over several orders of magnitude, it is necessary that $\Phi\gtrsim 10^{-4}$
ergs cm$^{-2}$ in order to produce a sufficient neutrino fluence for detection,
given that the $\nu_\mu$ fluence will generally be smaller than the
photon fluence even in the optimistic case that the photon radiation
originates substantially from hadronic processes.

\begin{figure}[t]
\vskip-0.3in
\center
\vspace*{15.0mm} % 
\includegraphics[width=11.0cm]{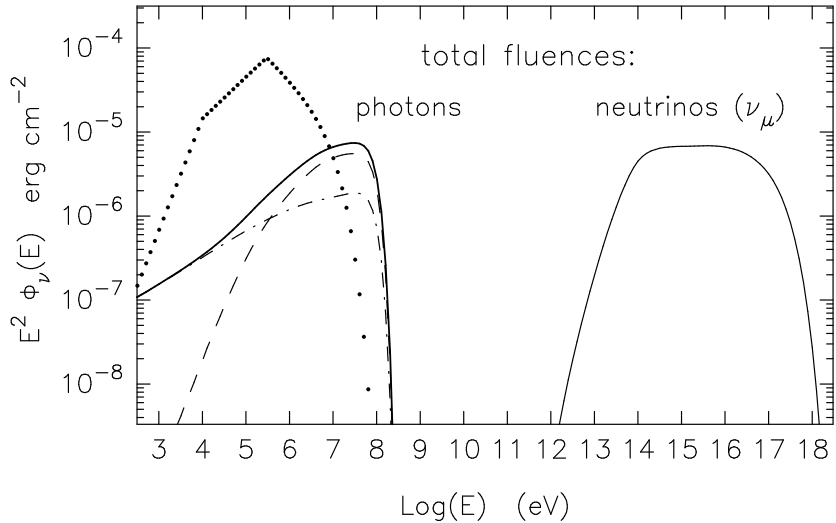}
%fig1_3.eps} 
%\vskip-1.0in
\caption{Differential energy fluence of photons and photomeson muon neutrinos for 
a collapsar-model GRB with hard X-ray fluence $\Phi_{tot} = 3\times
10^{-4}$ erg cm$^{-2}$ and $\delta = 100$. The dotted curve shows the
fluence of a GRB used for calculations, and the dashed and dot-dashed
curves show the Compton and synchrotron contributions to the photon
fluence from the electromagnetic cascade initiated by secondaries from
photomeson processes, respectively.  }
\label{fig2}
\end{figure}

In calculations presented here to illustrate the astrophysical
importance of photomeson production and subsequent electromagnetic
cascades, the observed synchrotron spectral flux in the prompt phase
of the burst is parameterized, as shown in Fig.\ 2,  by the expression $F(\nu) \propto
\nu^{-1} (\nu/\nu_{br})^{\alpha}$, where $h\nu_{br}=300$ keV,
$\alpha = -0.5$ above $\nu_{br}$ and an exponential cutoff
at 10 MeV, and $\alpha = 0.5$ when $10\, {\rm
keV} \leq h\nu \leq h \nu_{br}$. At lower energies, $\alpha= 4/3$.
The observed total hard X-ray (keV -- MeV) photon fluence $
\Phi_{tot} \cong t_{dur}\int_0^\infty d\nu F(\nu )$, where $t_{dur}$
is the characteristic duration of the GRB. We consider a source at
redshift $z=1$ and  assume
the hard X-ray fluence
$\Phi_{tot} \gtrsim 3\times 10^{-4} \,\rm erg \; cm^{-2}$.  One or two GRBs 
should occur each year above this fluence level. Here we take $s = 2$.

A total amount of energy $E^\prime = 4\pi d_L^2 f_{CR} \Phi_{tot}
\delta^{-3} (1+z)^{-1}$ is injected in the form of accelerated proton
energy into the comoving frame of the GRB blast wave. Here $z$ is the
redshift, $d_L$ is the luminosity distance, and
\begin{equation}
\delta = {1\over \Gamma(1-\beta\mu)}\;
\label{dop}
\end{equation}
is the Doppler factor. The factor $f_{CR}$ is the baryon-loading
factor, which gives the ratio of energy deposited in nonthermal
hadrons compared to the energy detected as electromagnetic radiation,
which is assumed to be provided by nonthermal electrons.  The energy
deposited into each of $N_{sp}$ light-curve pulses (or spikes) is
therefore $E_{sp}^\prime = E^\prime/N_{sp}\,$ ergs.  We assume that
all the energy $E_{sp}^\prime$ is injected in the first half of the
time interval of the pulse (to ensure variability in the GRB light
curve), which effectively corresponds to a characteristic variability
time scale $t_{var} = t_{dur}/2N_{sp}$.  The proper width of the
radiating region forming the pulse is $\Delta R^\prime \cong t_{var}
c\delta/(1+z)$, from which the energy density of the synchrotron
radiation can be determined \cite{ad01}.  We set the GRB prompt
duration $t_{dur} = 100\,$s, and let $N_{sp} = 50$, corresponding to
$t_{var} = 1\,\rm s$.  

The magnetic field is determined by assuming
equipartition between the energy densities of the magnetic field and
the nonthermal electron energies inferred from the synchrotron radiation
spectrum. For the parameters considered here, the equipartition magnetic 
field  range from several 100 G to a few kG (depending on $\delta$). 
For fields $B \gg$ kG, energy losses of the pions and muons can introduce
a break in the $\nu_\mu$ spectrum at multi-PeV energies \cite{muc99a}, 
but we neglect that effect in 
our calculations. The constraint on maximum 
possible proton energy imposed by the condition that the gyroradius is
less than the source size is also imposed. In addition, relativistic expansion
of the blast wave causes adiabatic cooling of the particles, which limits
neutrino production for a duration set by the blast-wave shell light-crossing time.

\begin{figure}
\center
\includegraphics[scale=0.5]{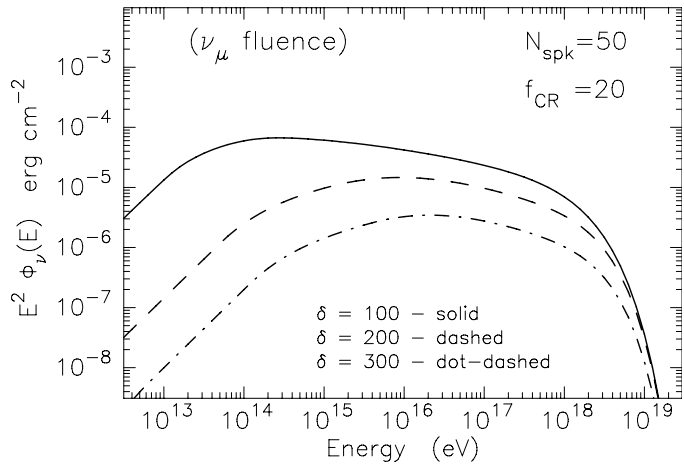}     
\caption{ The differential energy fluence of muon neutrinos calculated 
for a GRB at $z = 1$ with hard X-ray fluence of $3\times
10^{-4}$ ergs cm$^{-2}$ and Doppler factors $\delta = 100$, 200 and
300, and a nonthermal baryon-loading factor $f_{CR}= 20$.  }
\label{fig3}
\end{figure}

In Fig.\ 3 we show the neutrino fluence expected in the collapsar GRB
scenario from a model burst with photon fluence $\Phi_{rad} = 3\times
10^{-4}\,\rm erg\, cm^{-2}$ at redshift $z = 1$.  In order to demonstrate the
dependence of the neutrino fluxes on $\delta$, we consider 3 values of
$\delta$. The value of $f_{CR}$ is set equal to $20$ in this
calculation. Here we take $s = 2.2$.
The numbers of muon neutrinos that would be detected from a single
GRB with IceCube for these parameters and with
$\delta = 100,\, 200$ and 300 are $N_\nu = 1.32,\, 0.105 $ and 0.016,
respectively. We should note, however, that for the assumed value of
$f_{CR}$, the calculated total fluence of neutrinos (both $\nu_\mu$
and $\nu_e$) produced when $\delta = 100$ is $\Phi_{\nu ,tot} =7.2
\times 10^{-4}$ erg cm$^{-2}$, i.e., by a factor $7.2/3 = 2.4$ {\it
larger} than the assumed radiation fluence.  This means that the
maximum value of the baryon loading that could be allowed if the
high-energy radiation fluence is less than the X/$\gamma$ fluence for
this particular case should be about 8 -- 10, instead of 20, in order
that the hadronic cascade $\gamma$-ray flux is guaranteed not to
exceed the measured photon flux. Consequently, the expected number of
neutrinos for $\delta=100$ should be reduced to $\simeq 0.6$. On the
other hand, the neutrino fluence for the case $\delta =200 (300)$ is
equal to $\Phi_{\nu,tot} =1.4 \times 10^{-4} (3 \times 10^{-5}) \,\rm
erg\, cm^{-2}$, so this accommodates an increased baryon-loading from
$\lesssim 20$ up to $f_{CR}\simeq 45 (200)$, with the expected number
of neutrinos observed by IceCube being $N_{\nu ,corr} \simeq 0.23
(0.16)$.  If the radiation fluence at MeV -- GeV energies is allowed
to exceed the X/$\gamma$ fluence by an order of magnitude, a
possibility that GLAST will resolve, then the expected number of
detected neutrinos could be increased correspondingly.

For the large baryon load $f_{CR} \gtrsim 20$, which is required in
the hypothesis that GRBs are the sources of UHECRs, as discussed
in the next Section, calculations show
that 100 TeV -- 100 PeV neutrinos could be detected several times per
year from all GRBs with kilometer-scale neutrino detectors such as
IceCube \cite{da03,wda04}. 
It is important that at these energies the number of atmospheric 
background neutrinos expected for a km-scale neutrino detector in the time window
of a typical long GRB is negligible.
 Detection of even 1 or 2 neutrinos from
GRBs with IceCube or a northern hemisphere neutrino detector will
provide compelling support for this scenario for the origin of
high-energy and UHE cosmic rays.

\begin{figure}[t]
\vskip-0.3in
\vspace*{15.0mm} % 
\center
\includegraphics[width=8cm]{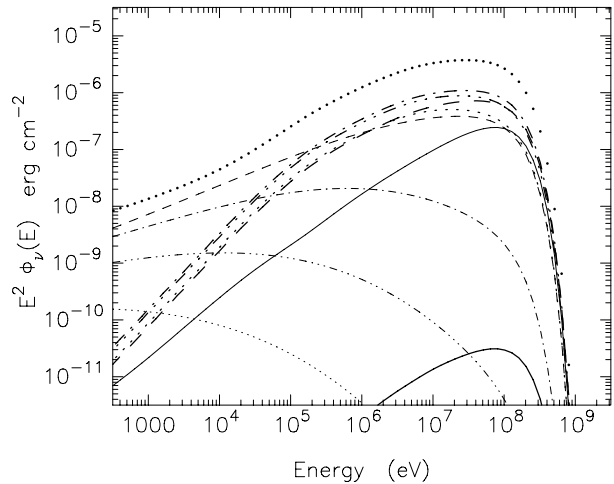}
%fig1_3.eps} 
%\vskip-1.0in
\caption{Differential photon energy fluence from an electromagnetic cascade initiated
by photopion secondaries in a model GRB, with parameters as given in
Fig.\ 2 and with $\delta = 100$. Five generations of Compton (heavy
curves) and synchrotron (light curves) are shown. The first through
fifth generations are given by solid, dashed, dot-dashed,
dot-triple--dashed, and dotted curves, respectively. The total cascade
radiation spectrum is given by the upper bold dotted curve. }
\label{fig4}
\end{figure}

A pair-photon cascade initiated by photohadronic processes between
high-energy hadrons accelerated in the GRB blast wave and the internal
synchrotron radiation field produces a $\gamma$-ray emission component
that appears during the prompt phase of a GRB, as shown in Fig.\ 4.
The various generations of synchrotron and Compton radiation initiated
by the cascade are shown, along with the total radiation spectrum. As
can be seen, the cascade radiation approaches the spectrum of an
electron distribution cooling by synchrotron losses, that is, a
spectrum with photon number index between $-1.5$ and $-2$.

Photomeson interactions in the relativistic blast wave also produce a
beam of UHE neutrons, as proposed for blazar jets \cite{ad03}, which
may escape from the site of the GRB to deposit UHECRs at distances ranging from 
the close exterior 
neighborhood of the blast wave to multi-parsec scales. 
Subsequent photopion production of these neutrons with photons outside the blast
wave can also produce a directed hyper-relativistic electron-positron
beam in the process of charged pion decay and the conversion of
high-energy photons formed in $\pi^0$ decay \cite{da04}. These
energetic leptons produce a synchrotron spectrum in the radiation
reaction-limited regime extending to $\gtrsim$ GeV energies, with
properties in the 1 -- 200 MeV range similar to that measured from GRB
941017 \cite{gon03}. Large fluence GRBs displaying these anomalous $\gamma$-ray
components are most likely to be detected as sources of high-energy
neutrinos \cite{gue04}.

Fig.\ 2 also shows calculations of photomeson neutrino production for a GRB
with $ \Phi_{tot} = 3\times 10^{-4}\,\rm erg \, cm^{-2}$ and $\delta =
100$, as well as the accompanying electromagnetic radiation induced by
pair-photon cascades from the secondary electrons and $\gamma$ rays
from the same photomeson interactions. The total number of $\nu_\mu$
expected with IceCube is $\cong 0.1$. The total fluence of cascade
photons shown here is contributed by lepton synchrotron (dot-dashed)
and Compton (dashed) emissions. For comparison, the dotted curve shows
the primary lepton synchrotron radiation spectrum assumed for the
calculations. The level of the fluence of the cascade photons is
$\approx 10$\% of the primary synchrotron radiation. This means that
the maximum allowed baryon loading for these parameters cannot exceed
a factor of $\approx 30$ in order not to overproduce the primary
synchrotron radiation fluence, as in the case of GRB 941017
\cite{gon03}. This limits the maximum number of $\nu_\mu$ to $\approx
3$ even in the case of large baryon loading for rare, powerful GRBs,
unless a very strong hadronic emission component accompanies the
GRB. This component, whose existence is indicated by joint analysis of
BATSE and EGRET/TASC data in GRB 941017 and at least two other GRBs \cite{gon03},
will be strongly detected by the GRB Monitor and Large Area Telescope
on GLAST.

\section{Cosmic Rays from GRBs}

The original argument \cite{vie95} that GRBs are the sources of UHECRs
was based on the coincidence between the energy density of UHECRs and
the amount of power necessary to supply cosmic rays with energies
$\gtrsim 10^{20}$ eV. These GZK cosmic rays, named after the
discoverers of the effect \cite{gzk}, are subject to strong photomeson
energy losses on cosmic microwave background photons.  The effective
distance for $10^{20}$ eV protons to lose 50\% of their energy is
$\approx 140$ Mpc \cite{sta00}, so that the photopion energy-loss timescale of
$10^{20}$ eV cosmic-ray protons is $t_{GZK} \cong 140$ Mpc/c $\cong
1.5\times 10^{16}$ s.

The energy density $u_{uhecr}$ of GZK cosmic rays observed near Earth,
as measured with the AGASA air shower array and the
High Resolution air fluorescence detector \cite{hires},
is $\approx 10^{-21}$ ergs cm$^{-3}$.  If these cosmic rays are
powered by GRBs with luminosity $L_{GRB}$ throughout the universe,
then
\begin{equation}
u_{uhecr} \cong \zeta\; {L_{GRB}t_{GZK}\over V_{prod}}\;
\label{u_UH}
\end{equation}
where $V_{prod}$ is the production volume of the universe. Here UHECRs
are assumed to be produced locally with an efficiency $\zeta$ compared
with the mean $\gamma$-ray power of GRBs.

The mean $\gamma$-ray fluence of BATSE GRBs is $F_\gamma \approx
3\times 10^{-6}$ ergs cm$^{-3}$ and their rate over the full sky is
$\dot N_{GRB}\approx 2/$day. If most GRBs are at redshift $\langle z
\rangle \sim 1$, as implied by redshift measurements of GRBs
detected with Beppo-SAX, which had similar triggering criteria as 
BATSE, then their mean distance is $\langle d \rangle
\approx 2\times 10^{28}$ cm. Thus the average isotropic energy
release of a typical GRB source is $\langle E_\gamma \rangle \approx
4\pi \langle d \rangle ^2F_\gamma/(1+z) \cong 8\times 10^{51}$ ergs,
implying a mean GRB power into the universe of $L_{GRB} \approx
2\times 10^{47}$ ergs s$^{-1}$. (This estimate is independent of the
beaming fraction, because a smaller beaming fraction implies a
proportionately larger number of sources.)  This implies that the energy
density observed locally is $u_{uhecr} \approx 10^{-22} \zeta$ ergs
cm$^{-3}$. 

Thus super-GZK particles could in principle be powered by GRBs if
roughly equal energies are deposited in super-GZK particles with
energies exceeding $10^{20}$ eV as is radiated by GRBs. Other effects,
for example, bolometric corrections to the total photon fluence, the
lower production rate of GRBs at $z \ll1$ than at $z \approx 1$, the
enhancement due to GRB-type sources which do not trigger GRB
detectors, complicate the estimate. 

In the model of Waxman and Bahcall \cite{wb99}, sources deposit UHECRs
with energies between $\approx 10^{19}$ eV and $10^{21}$ eV with a
$-2$ injection number spectrum.  In this way, the coincidence is preserved, but
the model requires a separate origin for UHECRs with energies below
the cosmic-ray ankle energy of $\lesssim 5\times 10^{18}$ eV.  In their model, the
ankle is interpreted as the energy of the transition between the
galactic and extragalactic cosmic rays.

By contrast, in the model of Ref.\ \cite{wda04}, cosmic rays with
energies between the knee at $\approx 3\times 10^{15}$ eV and the
second knee at $E_2\approx 5\times 10^{17}$ eV are mostly due to a single
or a few relatively recent Galactic GRB/supernova events that occurred
some $t_{0} \lesssim 10^{6}\,$yrs ago at distances $r_{0} \lesssim
1\,\rm kpc$ from us. UHECRs from extragalactic GRBs dominate at $E
\gtrsim E_2$. This model explains the entire CR spectrum from GeV up
to ultra-high energies (UHE) with a single population of sources,
namely SNe. 
The coincidence between GRB electromagnetic power and
UHECR power is however lost, and a large baryon load, $f_{CR} \gtrsim
20$ -- 50, is required in GRBs, as compared to $f_{CR}\approx 1$ in 
Ref.\ \cite{wb99}, due to the softer spectrum and greater injection-energy 
range. The transition from galactic to extragalactic component 
takes place at the second knee of the cosmic-ray spectrum, in accord with indications
for a change from heavy-to-light composition near the second knee \cite{bird}

A ``single-source" model was proposed earlier by Erlykin and
Wolfendale \cite{ew97}, who suggested that the knee could be due to a
single ``normal" supernova event that occurred some $t\sim 10^4\,\rm
yr$ ago within $r\sim 100\,\rm pc$ from us. This 
differs from the approach of Ref.\ \cite{wda04}, because the latter model
can explain acceleration of particles up to ultra-high energies by the
relativistic shocks formed by GRB outflows, which is very problematic
in the case of SNRs formed in the collapse to neutron stars.
Moreover, the much larger total energy of cosmic rays injected by 
a SN/GRB, which permits the
source to have occurred at larger ($\sim 1\,$kpc) distances and from a 
significantly older GRB than for a single normal SN source, makes it then 
easier to explain the likelihood of such an event, as well as the
low degree of anisotropy observed in HECRs.

The model of Ref.\ \cite{wda04} provides a way to explain the origin and the sharpness of
the knee at $E_1\simeq 3 \,\rm PeV$ as a consequence of pitch-angle
scattering of cosmic rays on plasma waves injected in the interstellar
medium (ISM) through dissipation of bulk kinetic energy of SNRs 
on the pc-scale Sedov length. Furthermore, 
the origin of the second knee in the cosmic-ray spectrum at $E_2 \simeq
4\times 10^{17}\,\rm eV$ is explained as a consequence
of diffusive propagation through scattering with turbulence injected on
a scale of $\sim 100\,\rm pc$, corresponding to the thickness
of the Galactic disk, which perhaps represents the largest natural
scale for effective injection of plasma turbulence in the Galaxy.  The
transition from Galactic to extragalactic CRs occurs around and above
the second knee.

\begin{figure}
\center
\includegraphics[width=7cm]{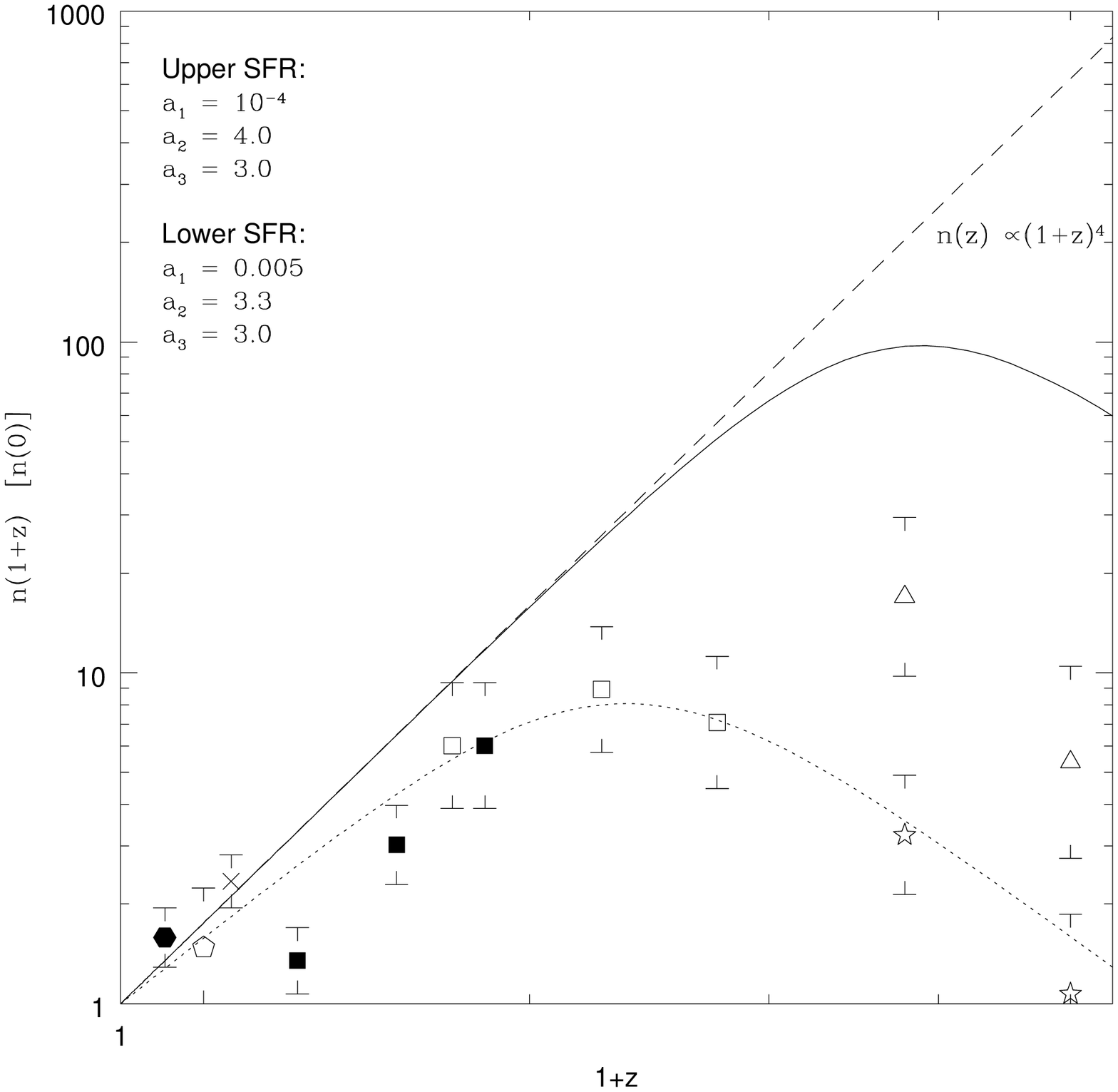}\hskip1.0cm
\includegraphics[width=7cm]{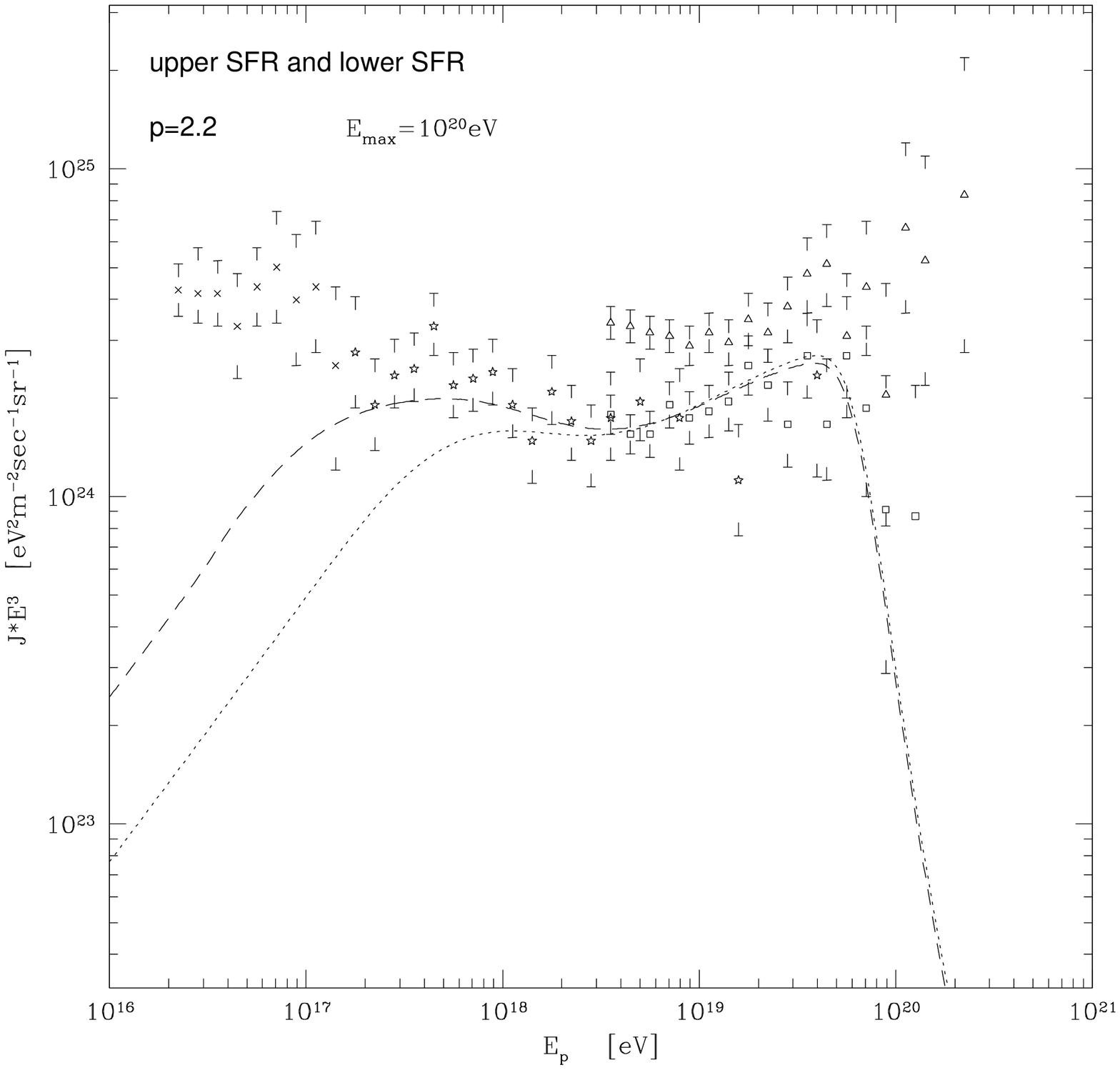}
\caption{{\protect \bf (a)} ({\it left}) 
The history of evolution of the star formation rate (SFR) in the universe 
as a function of redshift $1+z$, normalized to the current SFR.  
The dotted curve shows the lower limit to the SFR evolution implied
by measurements of the blue and UV energy density, and the solid 
curve shows the SFR corrected for dust extinction 
(see Ref.\ \cite{wda04} for detailed discussion). 
The dashed line displays the relation
$n(z)=n(0)(1+z)^4$ used by \cite{bgg04} for calculations of
the fluxes of extragalactic CRs. 
{\protect \bf (b)} ({\it right}) Calculated fluxes of extragalactic CRs 
assuming that the injection of UHECRs in the universe was due to 
GRBs with a rate density proportional to the 
 minimum (dotted curve) and maximum (solid curve) SFR functions 
shown in Fig.\ 5a. 
Note that the spectra are not normalized to each other at high
energies. Instead, the normalization for both of them corresponds to
the same value for the current ($z=0$) injection rate.    
}
\end{figure}

The rapid decline of the CR flux from local GRBs above the second knee results in
the  contribution of the extragalactic component to the all-particle spectrum
dominating near and above the second knee. Calculations of the extragalactic component 
are shown in Fig.\ 5 for our model of CRs from GRBs, which includes photomeson interactions, 
$e^+-e^-$ pair production, and adiabatic cooling of UHECRs \cite{bg88,ber90} during transport. 
We assume that the rate density of GRBs
is proportional to the cosmological SFR of the universe. 
For the two rates shown in Fig.\ 5a that correspond to minimum and 
maximum SFRs, calculations in \cite{wda04} result in the two spectra for the 
extragalactic component shown in Fig.\ 5b. An interesting result here is
that in the framework of this model, the ankle in the spectrum of CRs observed
at $E\simeq 3\times 10^{18}\,\rm eV$ is formed in the process of cooling of
UHE protons on cosmological timescales.

Similar spectral behavior for the extragalactic CR component at 
$E\geq 10^{18}\,\rm eV$ as shown in Fig.\ 5b,
where the ankle is explained
as a consequence of photopair losses of UHECRs formed 
at high redshift, 
was proposed also by Berezinsky and collaborators
\cite{bgg04}. They consider a model
where UHECRs are accelerated by active galactic nuclei
and assume cosmological evolution of the injection rate of UHECRs  
$\propto (1+z)^4$ (see Fig.\ 5a).
It remains to be studied if these two principal 
options (GRBs and AGNs) for the sources of UHECRs in the universe can be distinguished
from each other observationally as a result of differences in their
evolutionary histories. 

The spectra of UHECRs resulting from injection of UHECR protons in the universe
on cosmological timescales show a sharp (``GZK") cutoff above the GZK energy
$E\simeq 6\times 10^{19}\,\rm eV$. The UHECR spectrum in Fig.\ 5b 
agrees with the HiRes data, but is in disagreement with the AGASA results at 
$E\gtrsim 10^{20}\,\rm eV$. If Auger observations show any significant 
excess over the exponential GZK cutoff at these energies, this would
imply that there are other recent ($\lesssim 10^{8}\,\rm yr$) local
source/sources of extragalactic origin in our vicinity at
$\lesssim 10\,\rm Mpc$ that produce this flux. One 
possibility is that the excess would be due to cosmic ray
ions \cite{all05}.

Extragalactic sources could be connected
with starburst galaxies in the local group, such as M82 and NGC 253,
both at distances $r \sim 3.5\,\rm Mpc$. Taking into account that
the supernova rate in these galaxies is about 0.3 -- 1 per year, and that
the estimated GRB rate in our Galaxy is about (0.3 -- 1)\% of the 
supernova rate, the mean GRB rate in the starburst galaxies is 
estimated as one per $\sim 300$ -- 1000 yrs. If  
the total energy of CRs accelerated by a typical GRB is indeed about 
$10^{52}\,\rm ergs$, as for our local Galactic GRB, the characteristic
injection power of UHECRs from starburst galaxies averaged over the
timescale of $\sim 10^8$\,yr can be $\sim (1-3) \times 10^{42}\,\rm 
ergs\,s^{-1}$.

\section{GRBs in the Local Universe}

There are two important issues relevant to the question of 
the frequency and effects of GRBs in the Milky Way and the local
universe. These are (1) the rate of GRBs as inferred
from the GRB beaming factors; and (2) the dependence of GRBs
on the SFR of the universe and 
the tendency of GRBs to be found in sites of different metallicity. 
These questions are crucial to determine at what rate GRBs
occur in the Milky Way and whether they can be sources of cosmic
rays above the knee of the cosmic ray spectrum. At the present
time, it is not possible to answer either of these questions satisfactorily.

Regarding issue (1), achromatic beaming breaks in the afterglow
light curves of GRBs can be used to infer the opening angle of the 
GRB jets. Direct observations \cite{fra01} indicate that the mean measured 
opening angle is $\approx 0.1$ radian, so that the opening angle
of the GRB jets is $\approx 1/500^{th}$ of the full sky, leading to a rate
of GRBs in the Milky Way Galaxy of 1 every 3,000 -- 30,000 yrs \cite{der02}.
A self-consistent calculation \cite{gue} for both the redshift and opening 
angle
distribution suggests that the mean GRB opening angle is $\approx 1/75^{th}$
of the full sky, greatly reducing the rate in the Milky Way. The  initial 
(prompt)
opening angle
of the GRB jet may, however, be smaller than the opening angles of the lower bulk-Lorentz 
factor outflows that produce, at later times, 
beaming breaks observed at optical frequencies, as in structured 
jet models (e.g., \cite{gk06}). Hence the actual opening angles of the 
high Lorentz factor jets
making the $\gamma$ rays that trigger GRB detectors, and consequently
define the rate of GRBs in the Milky Way, remains very uncertain. 

The second important issue is whether GRBs follow the SFR
of the universe, as shown in Fig.\ 5b, or follow a significantly different 
rate. If GRBs are the population of massive stars that collapse to black
holes, then they may sample only a select portion of the high-mass stars that
produce the blue/UV luminosity used to determine the SFR. If the dependence
of the GRB rate on the SFR
changes with cosmic time, for example, due to 
the lower metallicities in the 
early universe, then the GRB rate in the recent epoch in the Milky Way may
be much lower than inferred from the high-redshift GRBs \cite{fru06,sta06}. 
The tendency of GRBs to favor low-metallicity hosts is not certain because
GRBs are preferentially found in OB associations \cite{fru06}, which may tend 
to have 
higher metallicity than the host galaxy;
 the host galaxy of GRB 980425 associated with SN 1998bw is
a spiral galaxy with fairly high metallicity \cite{sta06}; and high
metallicity is found for GRB 060206 at redshift 4.048 from absorption line
studies \cite{fyn06}. Moreover, the conclusions of \cite{sta06} rest
on only four GRBs with anomalously low beaming-corrected energies, 
which may simply indicate that underluminous GRBs have a tendency
to be born in low metallicity environments.

Evidence for local GRBs can be established by identifying
remnants of GRB in the Milky Way \cite{lp98}. 
The identification of a $\sim\! 1\,\rm Myr$ old GRB remnant (GRBR)
that happened at a $\lesssim 1\,\rm kpc$ from us, which explains
\cite{wda04} the origin of the knee(s) in the CR spectrum, is virtually 
impossible by now. However, there are other hopes to distinguish the 
radiation characteristics of the Galactic GRBRs from the radiation from
the remnants of ``ordinary" (nonrelativistic) supernovae in the 
Galaxy. 

The HESS 
(High Energy Stereoscopic System) collaboration has recently 
discovered a population consisting of several $\gamma$-ray 
sources in the Galactic 
plane \cite{science,Ginner}, which are spatially extended---up to tens of 
arcmin---and TeV-bright, 
but are quiet in all other frequency bands and remain unidentified.
The suppressed level of low-frequency synchrotron flux from 
TeV-bright sources is very unexpected for conventional SNRs.
It is proposed in Ref. \cite{abk06} through 
detailed calculations of particle diffusion and radiation processes 
and modeling of the spatial and spectral energy
fluxes detected, that at least one of these sources, 
HESS J1303-631 \cite{hess1303},
can be explained as a $\sim 10$-20 kyr old GRBR at a distance $\sim 10$-15
kpc from us. The hadronic origin of the detected TeV flux, implied by 
the non-detectable synchrotron flux, requires $\gtrsim 3\times 10^{51}$ ergs 
energy in relativistic protons with $E > 5\,\rm TeV$. Furthermore,
it is  qualitatively argued \cite{abk06} that the suppressed level of 
synchrotron flux could be the specific signature of the GRBRs. 
Unlike in the process of diffusive Fermi-type acceleration by 
nonrelativistic shocks for normal SNRs, the bulk kinetic energy of the 
relativistic shocks is converted to relativistic particles (mostly to
protons rather than to electrons, because of the large difference  
in the rest masses) at the very early stages  
while the shock is relativistic. This can explain the large baryon load
in the GRBs. Even at the early
non-relativistic stage of evolution of the GRBR, the magnetic fields
($\sim 0.1 \,\rm G$) are still high and can quickly remove the 
relativistic leptons from the pool of accelerated particles.
The most important observational feature predicted in \cite{abk06}
for the GRBRs, in contrast to SNRs,  is the very hard spectrum of 
$\gamma$ rays below 100 GeV energies.
A weak detection, confirming the hard spectrum,
or non-detection by GLAST of HESS J1303-631 
would therefore signify detection of the 
signatures of proton acceleration by relativistic shocks of a GRB.

Confirmation of HESS J1303-631 and perhaps some other extended unidentified 
TeV source as GRBRs would confirm the high rate, 
of order $10^{-4}\,\rm yr^{-1}$, of local GRBs. 
These questions are important for determining whether GRBs have
had an impact upon the evolution of life at Earth \cite{mel04,der02},  as
described in a related contribution to this focus issue \cite{tho06}.

\section{Summary and Conclusions}

The possibility that GRBs accelerate relativistic hadrons
could solve the problem of the origin of the high-energy cosmic rays. 
There are several ways to establish whether GRBs 
accelerate cosmic rays, from:  high-energy neutrino detection
from GRBs; anomalous radiation signatures
associated with hadronic acceleration and energy losses
in GRB blast waves; radiations made by neutrons that escape from 
the blast wave; and evidence for cosmic-ray production 
from GRB remnants in our own and nearby galaxies. 

In the case of neutrino and $\gamma$-ray production, 
at most only a few high-energy
$\nu_\mu$ can be detected with km-scale neutrino detectors even from bright
GRBs at the fluence level $\Phi_{tot} \gtrsim 3\times 10^{-4}
\,\rm erg \; cm^{-2}$, and only when the baryon loading is 
high \cite{da03,wda04}. This is because the detection of a single $\nu_\mu$
requires a $\nu_\mu$ fluence $ \gtrsim 10^{-4}
\,\rm erg \; cm^{-2}$ above 1 TeV. Since the energy release in high-energy
neutrinos and electromagnetic secondaries is about equal, this energy
will be reprocessed in the pair-photon cascade and emerge in the form
of observable radiation at $\gamma$-ray energies, and
this radiation cannot exceed the measured fluence in this regime. Detection
of high-energy neutrinos and hadronic $\gamma$-ray emission components
from GRBs will mean that GRBs have a large load in relativistic baryons, 
which is required for UHECR production.

The advent of the IceCube neutrino telescope at the South Pole and 
the launch of GLAST in late 2007 will provide the important instruments 
to establish whether GRBs accelerate ultra-relativistic hadrons. In the 
meantime, searches for GRB remnants at other wavelengths will indicate
the importance of GRBs in the local universes. With these developments, 
the longstanding puzzle of the origin of the cosmic rays---now nearly 
100 years since their discovery---should finally be answered.

\vskip0.2in 
\noindent We thank A.\ Reimer for permission to reproduce Fig.\ 1 from Ref.\ \cite{muc99},
and for useful comments on the manuscript.
The work of C.\ D.\ D.\ is supported by the Office of Naval Research. 
We also acknowledge support of GLAST Science Investigation DPR-S-1563-Y.

\end{document}